\documentclass[english,prl, reprint, footinbib]{revtex4-1}
\pdfoutput=1
\usepackage[T1]{fontenc}
\usepackage[latin9]{inputenc}
\usepackage{amsmath}
\usepackage{graphicx}

\makeatletter
 
 \@ifundefined{textcolor}{}
 {%
   \definecolor{BLACK}{gray}{0}
   \definecolor{WHITE}{gray}{1}
   \definecolor{RED}{rgb}{1,0,0}
   \definecolor{GREEN}{rgb}{0,1,0}
   \definecolor{BLUE}{rgb}{0,0,1}
   \definecolor{CYAN}{cmyk}{1,0,0,0}
   \definecolor{MAGENTA}{cmyk}{0,1,0,0}
   \definecolor{YELLOW}{cmyk}{0,0,1,0}
 }

\usepackage[colorlinks = true]{hyperref}
\usepackage{txfonts}
\usepackage[T1]{fontenc}
\usepackage{textcomp}

\makeatother

\usepackage{babel}

\begin{document}

\title{Localized microwave resonances in strained SrTiO$_{3}$ thin films}

\author{Patrick Irvin}

\affiliation{Department of Physics and Astronomy, University of Pittsburgh, Pittsburgh,
PA 15260}

\author{Jeremy Levy}

\email{jlevy@pitt.edu}

\affiliation{Department of Physics and Astronomy, University of Pittsburgh, Pittsburgh,
PA 15260}

\author{J. H. Haeni}

\affiliation{Department of Materials Science and Engineering, Pennsylvania State
University, University Park, PA 16802}

\author{Darrell G. Schlom}

\affiliation{Department of Materials Science and Engineering, Pennsylvania State
University, University Park, PA 16802}
\begin{abstract}
Local frequency-dependent polar dynamics of strained SrTiO$_{3}$
films grown on DyScO$_{3}$ are investigated using time-resolved confocal
scanning optical microscopy. Spectroscopic information is obtained
with $<1$ m spatial resolution over the frequency range 2-4 GHz.
Most of the DyScO$_{3}$ film is found to be spatially homogeneous,
in contrast to relaxed films. A strong correlation between spatial
and spectral homogeneity is revealed. In addition, resonant structures
are discovered that are localized both in space and in frequency.
\end{abstract}

\pacs{77.80.-e, 77.22.Gm, 85.50.-n}

\maketitle
Understanding the relationship between the polar structure and dynamic
response of ferroelectrics is critical for the development of integrated
devices. There are many factors which can produce dielectric dispersion
in these systems. Some are intrinsic to the phase transition itself,
while others depend on the existence of domain structures and their
dispersive properties. Arlt et al. have predicted that stripe domain
patterns in bulk BaTiO$_{3}$ single crystals will produce strong
dispersion in the GHz regime \citep{arlt_g_dielectric_1994}. Similarly,
polar complexes observed in relaxor ferroelectrics have been identified
by their characteristic frequency response \citep{viehland_deviation_1992,wakimoto_mode_2002}.
McNeal and coworkers have linked the domain state of BaTiO$_{3}$
(by way of grain and particle size) with microwave resonances \citep{mcneal_effect_1998}.
The relevant length scales for ferroelectrics span an unusually wide
range, from the atomic \citep{drezner_nanoferroelectric_2003} to
the crystal dimension itself \citep{pattnaik_new_1997}, and frequency
responses can also span from quasi-dc ($\sim1$ Hz) \citep{dunn_strain_2003}
to the ferroelectric soft mode ($\sim10^{11}$ Hz) \citep{vorobiev_microwave_2004,carlson_large_2000}.
Understanding how polar structure at a given length scale relates
to the dynamic response at a given frequency scale can shed light
on basic issues for device fabrication such as the fundamental limitations
for domain switching and mechanisms of microwave dielectric loss.

\begin{figure}
\begin{centering}
\includegraphics[width=88mm]{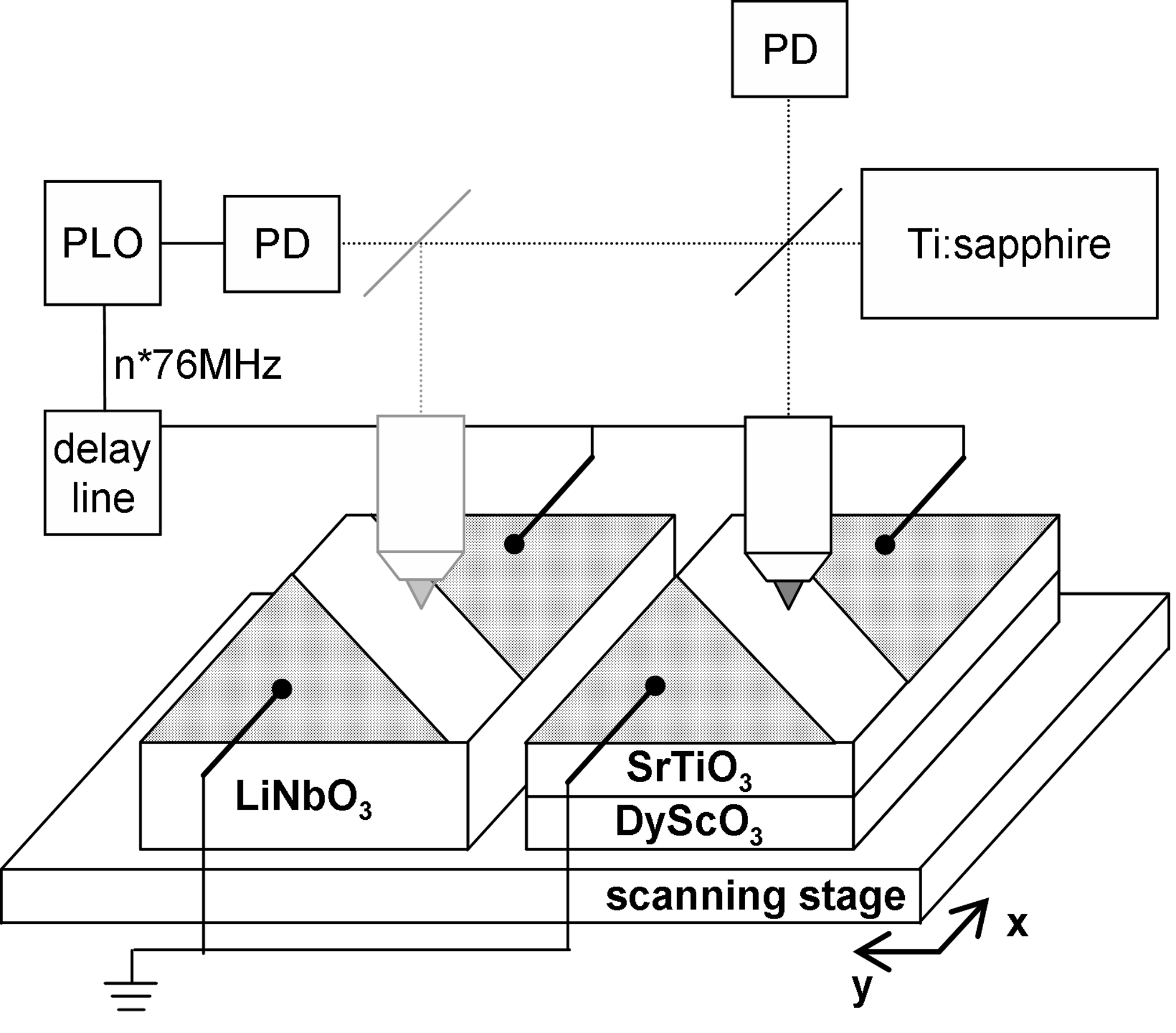}
\par\end{centering}

\caption{\label{fig:1}Diagram of experiment. Pulses from a mode-locked Ti:sapphire
laser are sampled by a fast photodiode (PD) that provides a reference
for a microwave-frequency phase-locked oscillator (PLO) whose frequency
is locked to a high harmonic of the fundamental repetition rate ($f{}_{1}=$
76 MHz) of the laser. This microwave signal is then applied to the
sample by using interdigitated silver electrodes. The microwave signal
at the sample is delayed in time relative to the incoming laser pulses
using a programmable delay line. The samples are raster scanned relative
to the microscope objective with a piezoelectric stage to produce
images. A single-crystal LiNbO$_{3}$ sample is used to produce a
reference phase at the different frequencies.}

\end{figure}

The dielectric constant of capacitors fabricated from ferroelectric
thin films can be changed with modest applied DC voltages, making
them suitable for tunable microwave devices such as phase shifters
\citep{mueller_ferroelectric_2001} and filters \citep{findikoglu_tunable_1996}.
A widely investigated material system is Ba$_{\text{x}}$Sr$_{\text{1-x}}$TiO$_{\text{3}}$,
whose Curie temperature $T_{c}$ can be varied from nearly 0 K for
pure SrTiO$_{3}$ ($x=0$) to 400 K for pure BaTiO$_{3}$ ($x=1$)
\citep{lancaster_thin-film_1998}. For thin films grown under most
conditions, non-uniform strain as well as stoichiometry fluctuations
can lead to an inhomogeneous broadening of $T_{c}$ over hundreds
of degrees. The growth of high-quality ferroelectric films is complicated
by a lack of suitably lattice matched substrates. Recently, bulk single
crystals of DyScO$_{3}$ have been synthesized and used to grow uniformly
strained SrTiO$_{3}$ films by molecular beam epitaxy \citep{haeni_rheed_2000,haeni_room-temperature_2004}.
The large biaxial tensile strain in this system results in ferroelectric
behavior at room temperature \citep{haeni_room-temperature_2004,chang_room-temperature_2004}.

Here we describe local spectroscopic investigations of polar dynamics
in a 500 � thick SrTiO$_{3}$ film grown on DyScO$_{3}$ \citep{haeni_room-temperature_2004},
We have developed an extension of time-resolved confocal scanning
optical microscopy (TRCSOM) that enables local polar dynamics to be
measured as a function of frequency as well as spatial location. Using
the electro-optic response to reveal polar dynamics in the SrTiO$_{3}$
film, we identify localized resonant features associated with the
periodic domain boundaries. While the domain structure appears to
be templated by the DyScO$_{3}$ substrate, the existence of these
resonant features represent the first direct experimental evidence
linking microwave resonances to domain structures.

A schematic of the experiment is shown in Figure \ref{fig:1}. An
ultrafast ($\sim120$ fs) mode-locked Ti:sapphire laser is used to
generate both the microwave electrical {}``pump'' field and the
optical probe pulse. The microwave pump signal is derived from a phase-locked
oscillator (PLO) that is locked to a high harmonic of the repetition
rate of the laser, $f{}_{1}=$ 76 MHz. The electrical signal is applied
to the SrTiO$_{3}$ film using Ag interdigitated electrodes deposited
on the film surface (gap width $d=10$ \textmu m), and are oriented
parallel to the <100> STO direction. Details about sample preparation
can be found elsewhere \citep{haeni_rheed_2000,haeni_room-temperature_2004}.
The laser pulses, focused to a diffraction-limited spot using a microscope
objective ($NA=0.85$), probe the electro-optic response at a fixed
phase of the microwave signal. The relative phase between electrical
and optical signals is controlled using an electrical delay line.
The amplitude of the microwave field is modulated at a low frequency
($\sim1$ kHz) and the resulting electrooptic signal is detected using
an optical bridge and lock-in amplifier \citep{hubert_new_1999}.
The reflected polarization of the laser light probes the electrooptic
response. The temporal response provides direct information about
local polar contributions to the microwave permittivity of the film
\citep{hubert_confocal_1997}.

TRCSOM images are acquired by raster scanning the sample relative
to the laser spot. Images are taken at ten different time delays td
(step size $\delta t=50$ ps) for 27 different microwave frequencies
$f_{n}=nf_{1}$, $26\leq n\leq53$, spanning the range 2-4 GHz. An
entire scan of frequencies, time delays, and spatial locations takes
approximately 12 hours to complete; thermal control of the TRCSOM
apparatus is maintained to within $T\sim0.02$ K in order to stabilize
the images sufficiently. The experiment is preformed at room temperature
(295 K), which is above $T{}_{c}$ in this sample. Post-processing
of the images is also performed to account for residual drift over
the acquisition period.

\begin{figure}
\begin{centering}
\includegraphics[width=88mm]{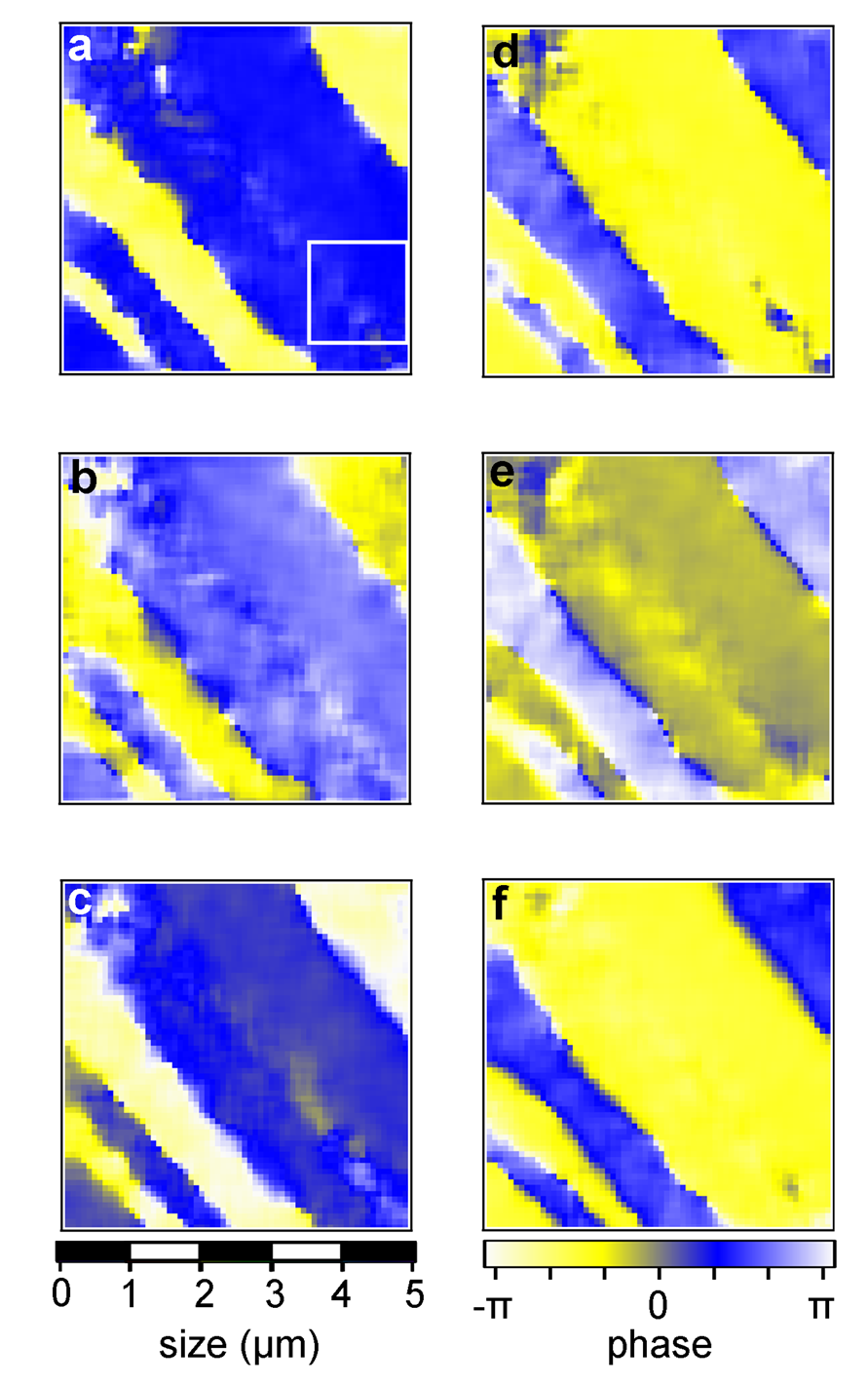}
\par\end{centering}

\caption{\label{fig:2}(Color online) Phase of ferroelectric response $\phi$
plotted as a function of location in the sample for six microwave
driving frequencies. (a) 2.2 GHz, (b) 2.43 GHz, (c) 2.74 GHz, (d)
2.96 GHz, (e) 3.27 GHz, (f) 3.5 GHz.}

\end{figure}

There is no intrinsic method for defining the absolute phase of the
incident microwave field relative to the optical probe, and the measured
phase changes in an uncontrolled way from one microwave frequency
to another. To produce a stable reference phase, TRCSOM measurements
were taken under identical conditions are taken on a single-crystal
LiNbO$_{3}$ reference sample, located several mm away from and connected
in parallel with the SrTiO$_{3}$ film. The phase of the linear electro-optic
response of the LiNbO$_{3}$, assumed to be constant over the frequency
range explored, is used to define a reference phase for the frequency-dependent
SrTiO$_{3}$ measurements.

The polar response of the SrTiO$_{3}$ film is well described by Fourier
components at angular driving frequency $\omega_{n}=2\pi f_{n}$and
second harmonic $2\omega_{n}$ \citep{hubert_mesoscopic_2000}:\begin{multline*}
S(t)=S_{0}+F_{1}\cos(\omega_{n}t)+F_{2}\sin(\omega_{n}t)\\
+P_{1}\cos(2\omega_{n}t)+P_{2}\sin(2\omega_{n}t)\end{multline*}

At each driving frequency, the sequence of images at various time
delays is used to produce a fit to Eq. 1 at each spatial location.
The result is an image of each of the four Fourier coefficients, \{$F{}_{1}$,
$F{}_{2}$, $P{}_{1}$,$P{}_{2}$\}. This analysis is performed for
each of the 27 discrete frequencies investigated.

\begin{figure}
\begin{centering}
\includegraphics[width=88mm]{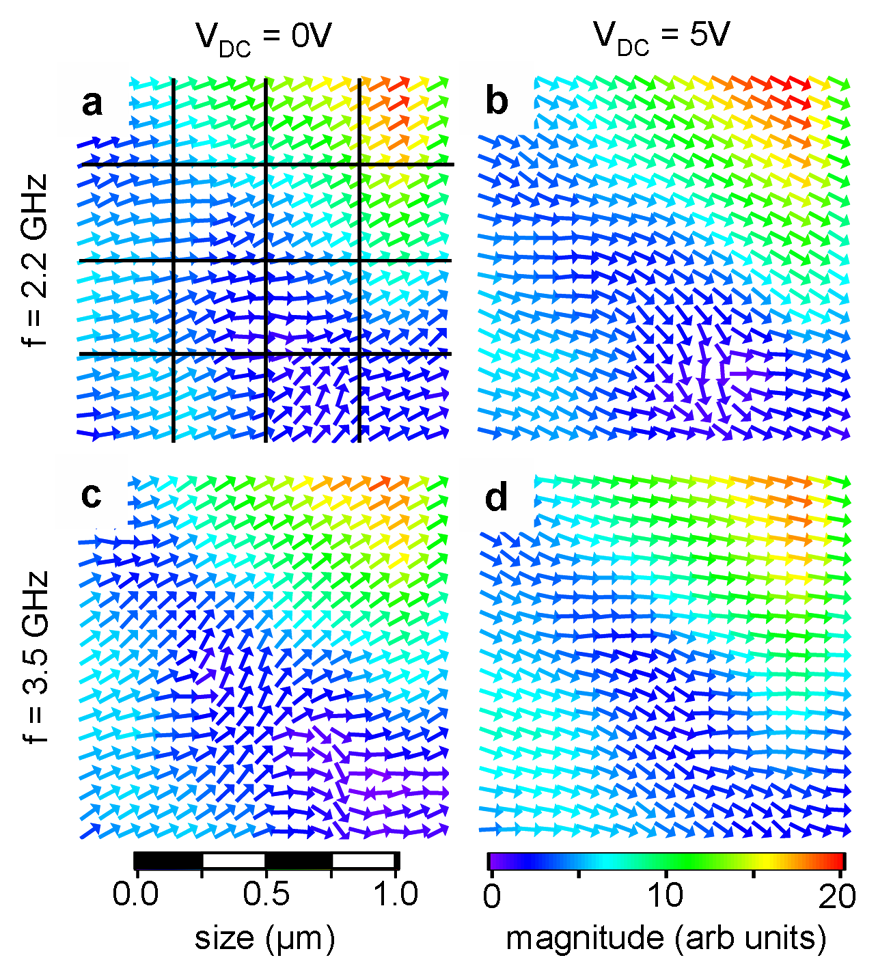}
\par\end{centering}

\caption{\label{fig:3}(color online) Vector plots showing magnitude (color)
and phase (angle) in the region identified in Fig. 2 (a). (a) and
(b) compare DC biases of 0 V and 5 V, respectively, at 2.2 GHz. (c)
and (d) compare DC biases at 3.5 GHz}

\end{figure}

Figure \ref{fig:2} shows images of the phase at six representative
driving frequencies. The large uniform regions visible in Figure \ref{fig:2}
are characteristic of the high quality of the SrTiO$_{3}$ film, and
are observed only with uniformly strained samples grown on DyScO$_{3}$
substrates \citep{haeni_room-temperature_2004}. The stripes correspond
to regions that are responding uniformly over the entire frequency
range investigated. Alternating stripes differ in phase by approximately
$\pi$, which is consistent with the existence of a domain wall boundary
separating them. These domain boundary regions exhibit a microwave
response that is much less uniform, and which exhibit dispersive behavior
that is localized in space and in frequency.

To illustrate the ferroelectric response within domains and near the
domain walls, we analyze subsections from the datasets shown in the
boxed region in Figure \ref{fig:2}(a). Figure \ref{fig:3} shows
vector field plots of the linear electro-optic response at two microwave
frequencies and two different dc bias voltages. Arrows are colored
according to the magnitude of the response, while their direction
indicates the local phase relative to the LiNbO$_{3}$ single crystal.
Regions of the sample that are far from the domain boundaries have
a uniform response, irrespective of applied frequency or DC bias.
However, select regions that are closer to the domain boundaries show
significant local dispersion when a DC bias is applied.

\begin{figure}
\begin{centering}
\includegraphics[width=88mm]{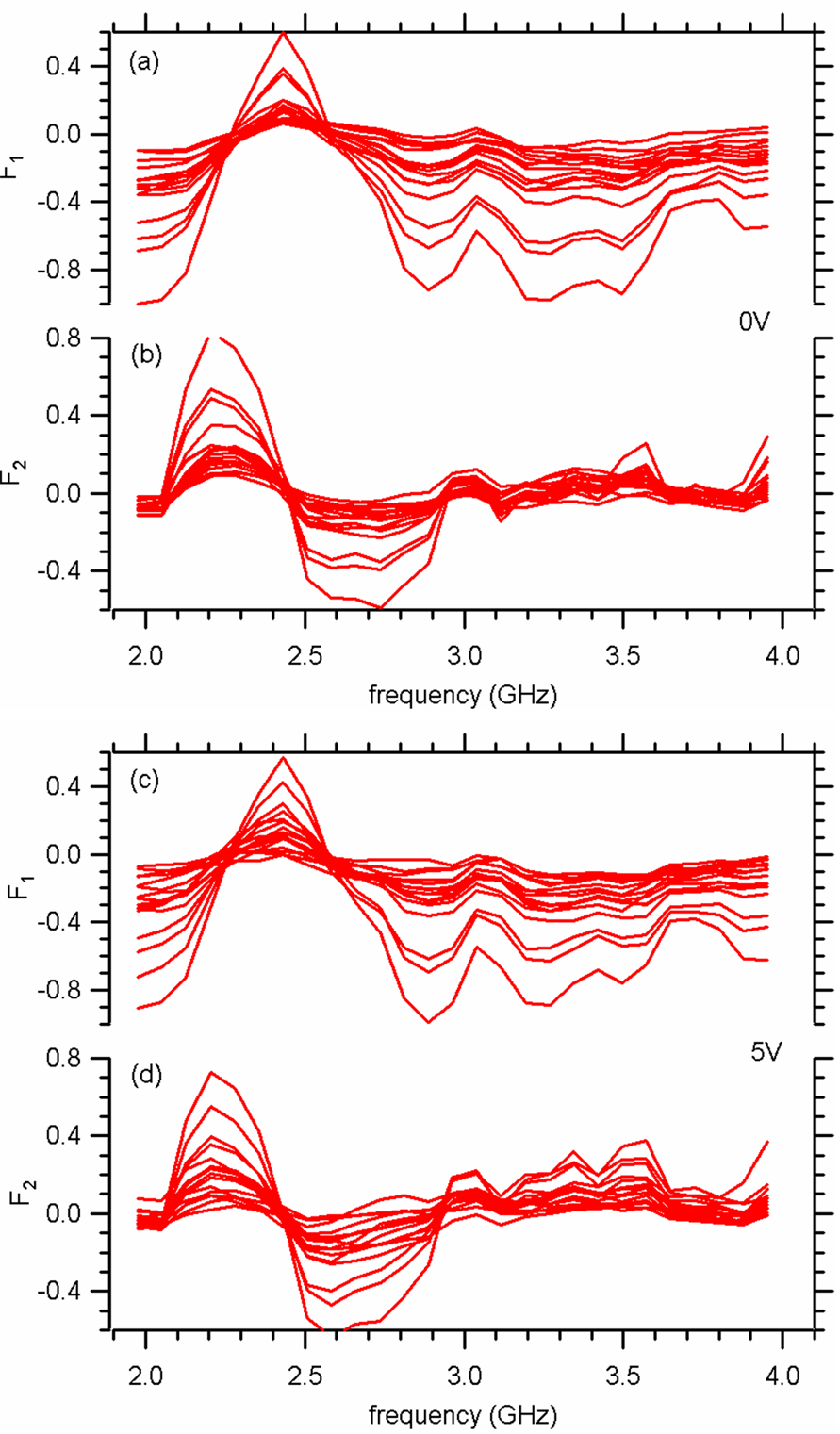}
\par\end{centering}

\caption{\label{fig:4}In- and out-of-phase components of ferroelectric response
plotted as a function of driving field frequency. Curves are taken
from the area shown in Fig. 3 and each line is an average over each
of the 16 sub-regions identified in Fig. 3(a).}

\end{figure}

To further investigate the local dynamics, the complex electro-optic
response $\mathbf{F}=F_{1}+iF_{2}$ is compared for several regions
of the sample. In Figure \ref{fig:4}, the linear electro-optic response
for the 16 sub-regions identified in Figure \ref{fig:3} are averaged
and plotted as a function of applied electric field frequency. Highly
dispersive responses are observed at 2.4 GHz and 3.5 GHz. When a DC
bias is applied, the dispersion increases at 2.4 GHz while decreasing
somewhat at 3.5 GHz.

A typical characteristic of relaxor ferroelectrics is their dielectric
dispersion characteristics, many of which can be understood by sound
emission due to domain walls vibration \citep{arlt_emission_1993}.
Biegalski \emph{et al} have shown that these SrTiO$_{3}$/DyScO$_{3}$
films show relaxor behavior in this frequency range \citep{biegalski}.
Additionally, the periodic domain structures observed by TRCSOM may
produce shear waves that interfere constructively or destructively,
depending on the driving frequency \citep{arlt_g_dielectric_1994}.
In addition to providing evidence of uniform ferroelectric response,
the stripe domain pattern we observe could also be the source of the
resonances we detect near the domain boundaries and within the domains
themselves.
\begin{acknowledgments}
We thank Stephen Kirchoefer for assistance in depositing the interdigitated
electrodes. Support from the National Science Foundation (NSF-0333192
and DMR-0103354) and the US Department of Energy is gratefully acknowledged.
\end{acknowledgments}
\bibliographystyle{apsrev4-1}
\bibliography{APL_STODSC_abbrev}

\end{document}